\documentclass[10pt,conference]{IEEEtran}

\usepackage{graphicx}
\usepackage{xcolor,soul}

\definecolor{codeColor}{RGB}{139,26,26}

\renewcommand{\hl}[1]{#1}

\newcommand{\citeme}[1]{%
  \begingroup
  \definecolor{hlcolor}{RGB}{255, 226, 176}\sethlcolor{hlcolor}%
  [\textcolor{orange}{\hl{\textbf{CITE}}}]%
  \endgroup
}

\usepackage{tikz}

\newcommand*\squared[1]{\tikz[baseline=(char.base)]{
\node[shape=rectangle,font=\bfseries,thin,draw=black,fill=yellow,text=black,inner sep=1pt] (char) {#1};}}

\graphicspath{{figures/}}

\def\BibTeX{{\rm B\kern-.05em{\sc i\kern-.025em b}\kern-.08em
    T\kern-.1667em\lower.7ex\hbox{E}\kern-.125emX}}
\begin{document}

\title{CONSTRUCT: A Program Synthesis Approach for Reconstructing Control Algorithms from Embedded System Binaries in Cyber-Physical Systems}


\author{\IEEEauthorblockN{
Ali Shokri\IEEEauthorrefmark{1},
Alexandre Perez\IEEEauthorrefmark{2},
Souma Chowdhury\IEEEauthorrefmark{3}, 
Chen Zeng\IEEEauthorrefmark{3}, 
Gerald Kaloor\IEEEauthorrefmark{2},
Ion Matei\IEEEauthorrefmark{2}, \\
Peter-Patel Schneider\IEEEauthorrefmark{2}, 
Akshith Gunasekaran\IEEEauthorrefmark{2}, and
Shantanu Rane\IEEEauthorrefmark{2}}
\IEEEauthorblockA{\IEEEauthorrefmark{1}Rochester Institute of Technology, NY, USA, as8308@rit.edu}
\IEEEauthorblockA{\IEEEauthorrefmark{2}Palo Alto Research Center (PARC), CA, USA, \{aperez, imatei, pfps, agunasekar, srane\}@parc.com}
\IEEEauthorblockA{\IEEEauthorrefmark{3} University at Buffalo, NY, USA, \{soumacho, czeng2, geraldka\}@buffalo.edu}
}

\maketitle

\begin{abstract}
We introduce a novel approach to automatically synthesize a mathematical representation of the control algorithms implemented in industrial cyber-physical systems (CPS), given the embedded system binary. The output model can be used by subject matter experts to assess the system's compliance with the expected behavior and for a variety of forensic applications. Our approach first performs static analysis on decompiled binary files of the controller to create a sketch of the mathematical representation. Then, we perform an evolutionary-based search to find the correct semantic for the created representation, i.e., the control law. We demonstrate the effectiveness of the introduced approach in practice via three case studies conducted on two real-life industrial CPS. 

\end{abstract}


\begin{figure*}[h]
  \centering
  \includegraphics[width=.97\textwidth]{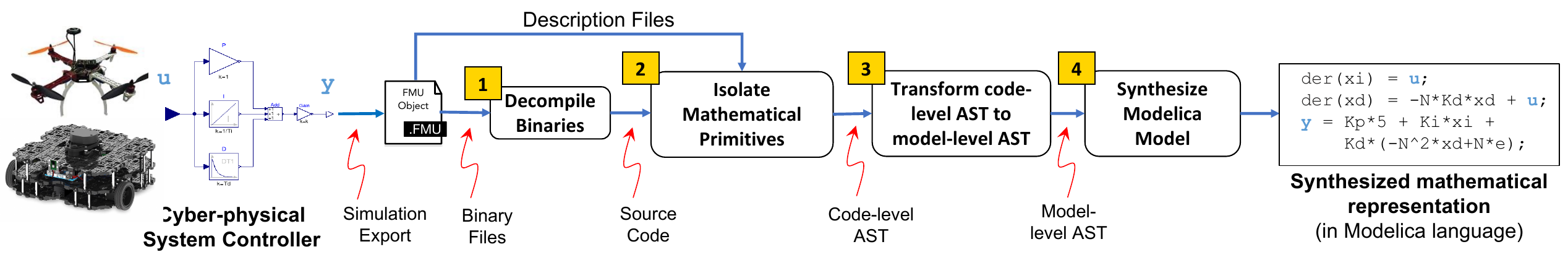}
    \vspace{-6pt}

  \caption{An overview of the \textsc{CONSTRUCT} approach.}
  \label{fig:CONSTRUCT}
          \vspace{-6pt}
\end{figure*}

\section{Background and the Problem}
\label{sec:Introduction}

Cyber-physical systems (CPS) consist of heterogeneous physical and computational components which combine to enable critical functionalities, e.g., providing energy to a city or flying an autonomous plane for delivering packages \cite{barrere2021analysing}. A key part of a CPS is the embedded software that implements the control algorithms and is in form of binaries.
For forensic investigations, especially in mission- and safety-critical industrial domains (e.g., military, energy, medical, transportation, and agriculture), it is crucial to reverse-engineer these algorithms as representation models to better understand their behavior and thus, investigate cause of failure in industrial accidents, detect counterfeit products or even infringement of intellectual property. 
The current techniques for construction of such representations heavily rely on intelligent experts (e.g., professional programmers) to manually review the decompiled binaries and create an abstraction of the software understandable by the code inspectors \cite{lyu2019safety}. This is a non-trivial, time-consuming, and error-prone task. 


To address these challenges, we introduce a novel approach for automatically synthesizing a mathematical representation of the controller software in a CPS. This representation enables subject matter experts to easily analyze the behavior, identify deviations from expected behavior, and avoid damaging threats.
The synthesized representation follows the syntax and semantic of Modelica modeling language~\cite{fritzson1998modelica}, which is widely used in industry and is easily understandable by subject matter experts. 
%
\section{Approach}
\label{sec:Construct}
Figure~\ref{fig:CONSTRUCT} shows an overview of our approach, called \textbf{\underline{C}ode-based M\underline{o}del Sy\underline{n}thesi\underline{s} Pla\underline{t}form fo\underline{r} re-Constr\underline{u}cting \underline{C}ontrol Algori\underline{t}hms} \textsc{\textbf{(CONSTRUCT)}}, which automatically constructs mathematical representations of controller algorithms from CPS binaries. The input to \textsc{CONSTRUCT} is a Functional Mock-up Unit (FMU) which is a file produced by CPS designers and contains the controller binaries as well as a textual description about the controller's I/O variables. The output of \textsc{CONSTRUCT} is a mathematical representation of the controller of CPS in Modelica language. The main idea is to decompile the given binary, locate controller-related instructions in the decompiled code, construct the \textbf{structure} of the mathematical representation, and finally, add the \textbf{semantic} to the created structure. Below, we provide more details on each step.

    \squared{1} \textbf{Decompile Binaries:} From description files in the FMU, \textsc{CONSTRUCT} identifies variable names and their attributes (e.g., type and initial value) that should be used in the mathematical model. It also decompiles the binary files inside the FMU using the popular Ghidra decompiler. The decompiled code contains symbolic (not actual) variable names. 
    
    \squared{2} \textbf{Isolate Mathematical Primitives:} CONSTRUCT performs static program analysis and localizes the mathematical primitives that are used in controller parts of the CPS. Some mathematical primitives differ in the decompiled version of the code compared to the original source code. CONSTRUCT incorporates a rule-based engine that is able to identify these complicated mathematical primitives in the decompiled code. 
    
    \squared{3} \textbf{Code-level AST to Model-level AST:} Next, CONSTRUCT creates ASTs of the isolated mathematical primitives in the decompiled binaries (C files) and translates them into algebraic equations in form of Modelica ASTs. Note that due to the decompilation of the binaries, the translated ASTs contain symbolic names, not the actual variable names used in the original source code of the controller.  
    
    \squared{4} \textbf{Modelica Model Synthesis:} \textsc{CONSTRUCT} uses Genetic Algorithm (GA) as an evolutionary search method to find the correct mapping between the symbolic names in the Modelica AST and the original I/O variable names retrieved from the description file inside the FMU. In this setting, the $i$th gene in a chromosome represents an I/O variable name that should be assigned to the $i$th symbol in the created Modelica AST. The baseline GA approach stochastically manipulates chromosomes and then tests whether the altered chromosome results in a syntactically and semantically correct Modelica AST (i.e., \textit{correct-by-testing (CbT)}). In contrast, we follow a \textbf{correct-by-construction (CbC)} paradigm where we carefully design GA operators (i.e., first population generation, mutation, and cross-over) such that every generated chromosome complies with Modelica syntax and semantic. This not only prunes the search space and reduces the number of trial and errors (and hence, makes the approach scalable), it also results in significantly more accurate mathematical representations. 
    To find the error of the generated mathematical representation, we provide the same time series input to the given binary as well as the synthesized mathematical model, find their output distance, and compute the mean squared error (MSE).
 
\section{Evaluation}
\label{sec:PreliminaryResult}
To showcase the performance of our approach, we present the results of our experimental studies on synthesizing mathematical representations from binaries of three different built-in controllers shipped with Modelica, namely, \textit{PI}, \textit{PID}, and \textit{LimPID}, that are used in two real-world industrial CPS: (1) a \textit{Turtlebot Waffle Pi} which is a four-wheeled robot used in ground-mission applications, e.g., surfing the area and building maps, and (2) a \textit{PX4 Quadcopter} which is an autonomous flying robot used in different industrial applications, e.g., package deliveries. The computational complexity of the embedded controllers (and thus, their to-be-synthesized mathematical representations) increases from PI controller where it has only proportional (P) and integral (I) blocks, to PID where there are proportional (P), integral (I), and derivative (D) blocks, and finally, LimPID where it has more complexity compared to the two previous controllers, e.g., has multiple inputs.
In our experiments, the population size for the GA was considered 400 and the maximum number of generations was 10.
%
\begin{figure}[ht]
  \centering
  \includegraphics[width=.47\textwidth]{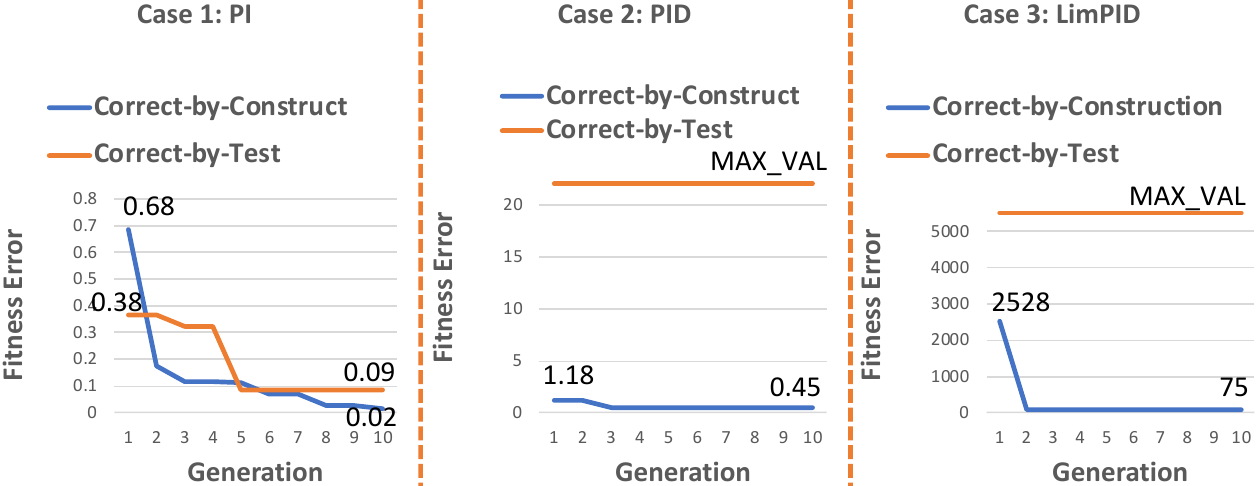}
  \caption{The result of synthesizing mathematical models based on the CbT (baseline) versus our CbC approach.}
  \label{fig:correct_by_construct_result}
\end{figure}
 Figure~\ref{fig:correct_by_construct_result} summarizes the results of our comparison study for our CbC approach against the baseline CbT approach. In this figure, X-axis denotes the generation number when we run the GA for finding the correct symbol to variable name mapping. Moreover, the Y-axis shows the MSE calculated based on the distance between the outputs of each of CbT and CbC approaches and the output of the actual binary, given the same input. 
 Even in the simplest controller (PI), our CbC approach is able to find more accurate mathematical representations compared to the baseline (CbT) approach. Interestingly, while in complex cases (PID and LimPID) the CbT approach is not able to generate even one representation that complies with Modelica language, not only our CbC approach can generate models that comply with Modelica syntax and semantic, but also the synthesized representations have relatively low error rates. The results show that the CbC approach outperforms the commonly used baseline approach, is able to generate accurate results to support subject matter experts in their investigations in industrial cases that the baseline CbT does not work, and is scalable to more complex controllers. 

\section{Conclusion and Future Work}
\label{sec:Conclusion}
In this paper, we introduced CONSTRUCT, a novel program synthesis approach that automatically creates mathematical representations of control algorithms in cyber-physical systems. This approach leverages a genetic algorithm for injecting the semantics into the created representation. In future work, we are interested in evaluating the utility of constraint-based solvers, such as SMT solvers, for making the model synthesis more efficient.

\bibliographystyle{IEEEtran}
\bibliography{bibliography}

\end{document}